\documentstyle[aps,prl,epsf,amsfonts]{revtex}
\def\ba{\begin{eqnarray}}
\def\ea{\end{eqnarray}}

\newcommand{\sect}[1]{\section{#1}\setcounter{equation}{0}}

\newcommand{\eqb}{\begin{equation}}
\newcommand{\eqe}{\end{equation}}
\newcommand{\dmb}{\begin{displaymath}}
\newcommand{\dme}{\end{displaymath}}
\newcommand{\pd}{\partial}

\newcommand{\eab}{\begin{eqnarray}}
\newcommand{\eae}{\end{eqnarray}}
\newcommand{\ra}{\right\rangle}
\newcommand{\la}{\left\langle}

\newcommand{\nc}{\newcommand}
\nc{\lab}{\label}
\nc{\eq}{Eq.\,(\ref}
\nc{\eqs}{Eqs.\,(\ref}
\nc{\tm}{\tiny\mbox}
\nc{\vivi}{very interesting and very important}
\nc{\al}{\alpha}
\nc{\ga}{\gamma}
\nc{\de}{\delta}
\nc{\ze}{\zeta}
\nc{\et}{\eta}

\nc{\Th}{\Theta}
\nc{\ka}{\kappa}
\nc{\lam}{\lambda}
\nc{\rh}{\rho}
\nc{\si}{\sigma}
\nc{\ta}{\tau}
\nc{\up}{\upsilon}
\nc{\ph}{\phi}
\nc{\ch}{\chi}
\nc{\ps}{\psi}
\nc{\om}{\omega}
\nc{\Ga}{\Gamma}
\nc{\De}{\Delta}
\nc{\La}{\Lambda}
\nc{\Si}{\Sigma}
\nc{\Up}{\Upsilon}
\nc{\Ph}{\Phi}
\nc{\Ps}{\Psi}
\nc{\Om}{\Omega}
\nc{\ptl}{\partial}
\nc{\del}{\nabla}
\nc{\be}{\begin{equation}}
\nc{\ee}{\end{equation}}
\nc{\bea}{\begin{eqnarray}}
\nc{\eea}{\end{eqnarray}}
\nc{\ov}{\overline}
\nc{\gsl}{\!\not}

\newcommand{\bi}[1]{\bibitem{#1}}
\newcommand{\fr}[2]{\frac{#1}{#2}}

\draft

\begin{document}

\title{(De-)Stabilization of an extra dimension due to a Casimir force}

\author{Ralf Hofmann$^1$, Panagiota Kanti$^2$, and Maxim Pospelov$^1$}
\address{$^1$Theoretical Physics Institute,  
         University of Minnesota,  
         Minneapolis, MN 55455, USA }

\address{$^2$Scuola Normale Superiore, Piazza dei Cavalieri 7,
I-56126 Pisa, Italy}
\maketitle

\centerline{ 
December 2000 }

\begin{abstract}
We study the stabilization of one spatial dimension in 
$(p+1+1)$-dimensional spacetime in the presence 
of $p$-dimensional brane(s), a bulk cosmological constant and 
the Casimir force generated by a conformally coupled scalar field. 
We find general static solutions to the metric which require the
fine-tuning of the inter-brane distance and the bulk cosmological
constant (leaving the two brane tensions as free parameters) corresponding to
a vanishing effective cosmological constant and a constant radion field. 
Taking these solutions as a background configuration, we perform 
a dimensional reduction and study the effective theory in the case 
of one- and two-brane configurations. We show that the 
radion field can have a positive mass squared, which corresponds to a 
stabilization of the extra dimension, only for a repulsive
nature of the Casimir force. This type of solution requires the presence of
a negative tension brane. The solutions with one or two positive
tension branes arising in this theory turn out to have negative 
radion mass squared, and therefore are not stable.

\end{abstract} 



\section{Introduction}

The past few years witnessed a growing interest among 
particle physicists and cosmologists toward models with extra space-like dimensions.
This interest was initiated by string theorists \cite{w}, who
exploited a moderately large size of an external 11th dimension in order
to 
reconcile the Planck and string/GUT scales. 
Taking this idea further, it was shown that 
large extra dimensions allow for a reduction of the fundamental
higher-dimensional 
gravitational scale down to the TeV-scale \cite{Dim}. 
An essential ingredient of such a 
scenario is the confinement of the standard model (SM) fields on field
theoretic defects, 
so that only gravity can access the large extra dimensions. 
These models are argued to make contact with an 
intricate phenomenology, with a variety of consequences for collider
searches, low-energy precision measurements, rare decays and
astroparticle 
physics and cosmology. However, the mechanisms, responsible for the 
stabilization of extra dimensions, remain unknown. The fact that 
the size of extra dimensions is large as compared to the 
fundamental scale also remains unexplained. An alternative solution to
the 
hierarchy problem was proposed in Ref.\,\cite{RS1}. 
This solution appeals to the possibility of a 
strongly curved extra dimension limited by two branes with positive and 
negative tensions, with the scale factor in the bulk space between the
two branes
changing by orders of magnitude within a distance of several Planck
lengths. 
In this case, the ratio of fundamental energy scales on the two branes 
is given by a large warping factor. However, the resolution of the hierarchy
problem is possible only in the case where the observable brane (on which
the SM fields are trapped) is the one with the negative tension. 

The early studies of cosmologies in the brane models revealed an unusual 
dependence of the Hubble parameter $H= \dot a / a $ on the energy density
of matter on the brane \cite{LOW,LK,BDL}. In fact, as it was shown in
Ref. \cite{BDL}, an abnormal behavior $H\sim \rho^{(4)}$ may persist in 
a later
epoch, {\em if} the stability of the extra dimension is achieved via a 
fine-tuned cancelation between positive and negative energy densities on 
the two branes. As it was first shown in Ref. \cite{kkop1}, 
a natural resolution to this problem comes from the stabilization of the
extra dimension, which also removes the necessity for an unphysical 
fine tuning between energy densities on different branes. 
The smooth transition to the 
four-dimensional cosmology and Newtonian 4-d interaction requires 
the presence of a non-vanishing (55)-component of the energy-momentum
tensor. 
It was subsequently shown in Refs. \cite{randall3,kkop2} that the value of
$T_{55}$ is automatically adjusted to a value proportional to 
$\rho^{(4)}-3p^{(4)}$ for a 
generic stabilization mechanism. It was also observed in 
\cite{randall3} that the stabilization mechanism is crucial for getting
a 
consistent solution to the gauge hierarchy problem on the negative
tension 
brane in the set up of Ref. \cite{RS1} (see also \cite{kim}). 
Subsequently, it was demonstrated that, in the presence of a 
stabilization mechanism, the solution to the hierarchy problem in the
two 
positive tension brane models is also possible \cite{kop1,kop2}. From a 
particle physics point of view, these models are more appealing 
than the models with negative tension branes, which 
cannot be realized as field theoretic solitonic solutions. 

Since the nature of the stabilization mechanism is yet unclear, it is 
reasonable to study a {\em minimal} possibility, which introduces one 
scalar field in the bulk. Classical stabilization forces due to
non-trivial 
background configurations of a scalar field along an extra dimension
were first discussed by Gell-Mann and Zwiebach \cite{GZ}.
With the revived interest in extra dimensions and brane worlds, a 
modified version of this mechanism, which exploits a classical 
force due to a bulk scalar field with 
different interactions with the brane(s), received significant 
attention \cite{Goldberger,group1,group2,Wisc}. However, as it was shown
in Refs.~\cite{kop1,Wisc}, a classical scalar interaction is not useful
for the stabilization of two positive tension branes. 

In this paper we study another stabilization mechanism due to 
a Casimir force generated by the quantum fluctuations about a constant
background 
of a conformally coupled massless scalar field. This effect was
initially studied in Ref.~\cite{Horava} in the context of M-theory, and
subsequently in Ref. \cite{Garriga} for a background Randall-Sundrum
geometry. The same effect was further investigated in \cite{more}.
Recently, S. Mukohyama \cite{Muko} 
and our group \cite{unpublished} found an 
exact static solution in 5 dimensions in 
the presence of the Casimir force and the bulk 
cosmological constant. 
By imposing boundary conditions on the branes, one may transfer the
breaking 
of the 5-d Poincar{\'e} invariance into the bulk, naturally 
generating an anisotropic energy-momentum tensor with $T_{5}^{5}$ different
from $T^0_0$ and $T^i_i$. 
It turns out that there are static solutions 
admitting one or two positive tension branes. 
The purpose of the present paper is to investigate the stability of 
these solutions. 

In the next section, we present
the theoretical framework of our analysis. We make an attempt to
derive time-dependent solutions in a $d$-dimensional, conformally-flat
spacetime, where there is a bulk cosmological constant and Casimir
stress. In the same
section we find an exact, static background in the case of positive,
negative or zero bulk cosmological constant and for Dirichlet or Neumann 
boundary conditions for the scalar field. 
We formulate the fine-tuning conditions which allow for such a solution
to exist. These conditions correspond to a vanishing effective
four-dimensional cosmological constant and a constant radion field.
In section 3, we study an effective field theory which is 
obtained after dimensional 
reduction in the presence of the static, gravitational background
determined in the case of a negative bulk cosmological constant. 
The effective potential for the size of the extra dimension is
calculated and, depending on the nature of the Casimir force,
it is either positive or negative. 
Negative mass squared arises due to an 
attractive Casimir force and corresponds to an unstable extremum. 
Unfortunately, the models with single positive tension brane or 
two positive tension branes fall into this category. 
Positive $m^2$, i.e.
true stabilization, arises in the case of a repulsive Casimir force
and {\em requires} the presence of the negative tension brane.

\sect{One or two $p$-branes in $(p+2)$ dimensions}

\subsection{Energy-momentum tensor due to the Casimir effect 
and the equations of motion in $d$ dimensions}

We start from the classical 
expression for the $d$-dimensional action describing a gravitational
field, a conformally-coupled scalar field $\phi$, and two $p$-branes
embedded in $d=p+2$ dimensions.
The branes have tensions 
$\Lambda_{1}$ and $\Lambda_2$ and are located at $z_1<z_2$, respectively,
where $z$ denotes the coordinate 
along the compact extra dimension. The bulk between the 
branes is characterized by a nonzero 
cosmological constant $\Lambda_B$ of an arbitrary sign.
The action of the system is given by\footnote{Throughout this paper, we 
follow Wald's conventions \cite{wald}: The metric signature is
$\eta_{MN}=(-,+,...,+)$ and the Riemann tensor is defined as
$R^\sigma_{\,\,\rho\mu\nu}=\partial_\mu \Gamma^\sigma_{\rho\nu}-
\partial_\nu\Gamma^\sigma_{\rho\mu} + ...\ $.}

\eqb
\label{act}
S=-\int d^{d-1}x\int dz \sqrt{-g_{(d)}}\,\left(- \frac{1}
{2\kappa^2_d}R_{(d)}+\Lambda_1\,\delta(z-z_1)+\Lambda_2\,\delta(z-z_2)+
\La_B+\frac{1}{2}
\pd_M\phi\,\pd^M\phi+\gamma R_{(d)} \phi^2\right)\ ,
\eqe
where $M=0,...,d-1$, $x^0\equiv t$, $x^{(d-1)}\equiv z$, and 
$\gamma\equiv(d-2)/(4(d-1))$. We also assume a $Z_2$ 
symmetry represented by mirror transformations 
about the branes in the z coordinate. 

At the classical level the stress energy tensor in the bulk is
determined 
only by $\Lambda_B$ which makes it isotropic, $T^M_N = -\La_B\,\delta^M_N$.
At the quantum level the $d$-dimensional 
isotropy of  $T^M_N$ can be broken by the Casimir stress. 
Following Ref. \cite{Garriga}, we exploit the relation 
between the quantum induced part of the energy-momentum tensor in 
flat spacetime and in conformally flat 
geometries. The vacuum average $\la T_{MN}\ra$ in a 
free field theory bounded in a $d$-dimensional 
flat spacetime can be inferred from the effective 
potential of a constant classical background, 
where the fluctuations are to obey the boundary conditions, 
by symmetry considerations or 
directly from the corresponding propagator \cite{Casimir}. 

In the case, where the 
boundaries are two parallel $p$-branes, separated by a spatial 
distance $L$, the tracelessness of the (improved \cite{Callan}) 
energy-momentum tensor implies the following result
\eqb
\label{dL}
\la (T^M_N)\ra^f=\frac{\alpha}{L^d}\,
\mbox{diag}\left(1,1,\cdots,1,-(p+1)\right)\ ,
\eqe
where $\alpha$ is dimensionless and depends on the fields of the theory,
the 
boundary conditions, and $d$. The same form of $T^M_N$ will hold for
a single brane configuration, which can be obtained by identifying 
two branes of equal tensions. 

It is common wisdom \cite{Garriga,BD} that the 
Casimir effect $\la (T^M_N)\ra^{cf}\not=0$ 
due to a conformally coupled, massless field in
a conformally flat, odd-dimensional spacetime is related to the 
corresponding flat-space expression as follows
\eqb
\label{con}
\la (T^M_N)\ra^{cf}=a^{-d}\la (T^M_N)\ra^{f}\ ,
\eqe
where $a^2=a^2(x)>0$ denotes the conformal factor transforming 
the metric from flat to curved geometry. Note that there is 
no restriction to the 
dependence of $a$ on the spacetime 
coordinates $x=(t,x_1,\cdots,x_p,z)$ other 
than smoothness and definiteness. This conversion formula for the vacuum
averages of the 
energy-momentum tensor in curved and flat space 
can be obtained by an explicit calculation, using a 
field redefinition $\tilde \phi = a^{(d-2)/2}\phi$, and it 
is only valid for odd $d$ \cite{BD}. 

Our goal is to find whether this contribution to the total energy-momentum
tensor of the bulk leads to a true stabilization of the extra dimension.
To study that, we consider, for now, only a dependence of $a$ on $t$ and $z$:
\be
ds^2=a^2(t,z)\,(- dt^2+ dx_1^2 + dx_2^2 + ... + dx_p^2 + dz^2)\ .
\label{metric}
\ee
As for the field $\phi$ the background configuration is $\phi_c=const$. 
Varying the action of Eq.\,(\ref{act}) with respect to the metric, one
obtains 
the following Einstein's equations $G_{MN}= \kappa_d^2\, T_{MN}$
\ba
&~&  G_{tt}=\frac{p}{2}\,(p+1)\,\frac{\dot{a}^2}{a^2}-p\,\frac{a''}{a}
-\frac{p}{2}\,(p-3)\,\frac{a'^2}{a^2}
= -{\kappa}^2_d\,a^2\,\Bigl[-\Lambda_B+
\frac{\alpha}{(aL)^{p+2}}\Bigr]\,,\label{00}\\[4mm] 
&~&  G_{ii}=p\,\Bigl(\frac{a''}{a}-\,\frac{\ddot a}{a}\Bigr)
+ \frac{p}{2}\,(p-3)\,\Bigl(\frac{a'^2}{a^2}-\frac{\dot a^2}{a^2}\Bigl)= 
{\kappa}^2_d\,a^2\,\Bigl[-\Lambda_B+
\frac{\alpha}{(aL)^{p+2}}\Bigr]\,,
\label{ii}\\[4mm]
&~&  G_{tz}=p\,\Bigl(2\,\frac{a'}{a}\,\frac{\dot{a}}{a}
-\frac{\dot{a}'}{a}\Bigr)=0\,,
\label{05}\\[4mm]
&~&  G_{zz}=\frac{p}{2}\,(p+1)\,\frac{a'^2}{a^2} -p\,\frac{\ddot{a}}{a} 
-\frac{p}{2}\,(p-3)\,\frac{{\dot a}^2}{a^2}=
{\kappa}^2_d\,a^2\,\Bigl[-\Lambda_B-
\frac{(p+1)\,\alpha}{(aL)^{p+2}}\Bigr]\ .
\label{55}
\ea
Thereby, the dots and primes denote derivatives with respect to $t$ and
$z$, 
respectively.      

Keeping the brane coordinates fixed, we now try to find general,
time-dependent solutions of
Eqs.\,(\ref{00})-(\ref{55}). Taking the sum of (\ref{00}) and
(\ref{ii}), 
an ordinary differential equation can be derived for $a(t,z)$ with
respect to time. An 
easy integration yields the solution
\be
\label{time}
a(t,z)=\frac{1}{f(z)\,(t-t_0) + g(z)}\, , 
\ee
where $f(z)$ and $g(z)$ are associated with the initial data $a(t_0,z)$
and $\dot{a}(t_0,z)$. 
Substituting the solution (\ref{time}) into Eq.\,(\ref{05}), gives
$f(z)=c_0\equiv const$. With this information, insertion of (\ref{time})
into Eq.\,(\ref{55}) gives
\eqb
\label{cons}
\frac{p(p+1)}{2}\left(g^\prime(z)^2-
c_0^2\right)\nonumber\\ 
=-\kappa_d^2\left(\La_B+
\frac{(p+1)\alpha}{L^{p+2}}\,[c_0(t-t_0)+g(z)]^{p+2}\right)\ .
\eqe
The {\it rhs} of the above equation becomes time-independent, like
the {\it lhs}, only if $\alpha=0$ or $c_0=0$. 
Let us consider first the case
$\alpha=0$, i.e. disregard the Casimir stress. Upon integration, Eq.
(\ref{cons}) then 
provides the solution for the function $g(z)$, and the
warp factor takes the final form
\be 
\label{time1} 
a(t,z)=\frac{1}{c_0\,(t-t_0) + c_1\,z}\, , \qquad {\rm where} \quad
c_1=\pm \sqrt{c_0^2-\frac{2 \kappa^2_d}{p(p+1)}\,\Lambda_B}\,. 
\ee 
Note that the above bulk solution exists for every value, positive,
negative or zero, of the bulk cosmological constant. The singular distribution
of energy at the location of the two branes generates the so-called 
{\it jump} conditions~\cite{israel}, which, in conformal coordinates,
take the form
\be 
\left.\frac{[a^\prime]}{a^2}\right|_{z_i}=
-\frac{\kappa^2_d}{p}\,\La_i\ , \qquad (i=1,2)\ . 
\label{jump}
\ee
Substituting the solution (\ref{time1}) in the above conditions, we
arrive at the constraints
\be
c_1=\frac{\kappa^2_d}{2p}\,\Lambda_1=-\frac{\kappa^2_d}{2p}\,\Lambda_2\, ,
\ee
which reveals the extreme fine tuning between the positive and negative
self-energies of the two branes necessary for the stabilization of 
the inter-brane distance in the absence of a stabilizing potential,
even in the time-dependent case. The fine tuning which is now relaxed,
due to the coefficient $c_0$, is the one that relates the brane
self-energies to the bulk cosmological constant. As a result,
the effective 4-d cosmological constant does not vanish, thus, inducing
the assumed evolution in time. One can easily check that, in the limit
$c_0 \rightarrow 0$, the extreme fine-tuning between all of the parameters
of the theory reappears, i.e.
\be
\Lambda_1^2=\Lambda_2^2=\frac{8p}{(p+1)}\,\frac{|\Lambda_B|}{\kappa^2_d}\,.
\ee
In this limit, note that the solution exists only for negative bulk
cosmological constant and the warp factor assumes the form 
$a(z)=l/z$, where $l=\sqrt{p(p+1)/(2\,\kappa^2_d\,\Lambda_B)}$ is
the d-dimensional AdS radius. For the specific case of a 3-brane
embedded in a 5-dimensional spacetime we easily recover the solution 
of Ref.~\cite{RS1} as presented in Ref.~\cite{Garriga} in conformal coordinates.

If, instead of $\alpha=0$, we choose alternatively $c_0=0$,
we re-introduce the Casimir stress in our analysis but, at the same time,
we loose the time-dependence of our solution. In the presence of 
the Casimir stress, time-dependent
solutions most probably require a departure from 
conformal geometry which immensely complicates the whole problem.
However, 
it is possible to construct an effective theory exploiting 
the large separation of the energy scales: Casimir stress and 
$\Lambda_B$ are presumably determined in units of a 
large fundamental $d$-dimensional 
gravitational scale. On the contrary, the characteristic 
frequencies of time-dependent perturbations are very small 
as compared to this scale for any 
cosmological epoch starting from reheating. 
Therefore, it seems reasonable 
to find static solution(s) in $d$ dimensions to perform a 
dimensional reduction about. Studying small and ``soft'' perturbations 
around theses solutions, one can perform the 
stability analysis for the 
extra dimension in the resulting $(d-1)$-dimensional, effective field theory.

\subsection{Static solutions}

Assuming the presence of the Casimir stress in the bulk, and of a bulk
cosmological constant, 
we seek static solutions for the warp factor $a=a(z)$. The system of 
Eqs.\,(\ref{00})-(\ref{55}) then reduces to the following set of
independent equations
\ba
&~& p\,\frac{a''}{a}
+\frac{p}{2}\,(p-3)\,\frac{a'^2}{a^2}
= {\kappa^2_d} \, a^2\,\Bigl[-\Lambda_B+
\frac{\alpha}{(aL)^{p+2}}\Bigr]\,,\label{eq1}\\[3mm] 
&~& \frac{p}{2}\,(p+1)\,\frac{a'^2}{a^2}= {\kappa^2_d}\,a^2\,
\left[-\Lambda_B-\frac{(p+1)\alpha}{(aL)^{p+2}}\right]. \label{eq2}
\ea
As we will see these equations can be integrated explicitly. 
Multiplying (\ref{eq1}) by $(p+1)$ and adding (\ref{eq2}), 
one obtains
\be
\label{res}
p\,(p+1)\,\frac{a''}{a}
+\frac{p}{2}\,(p+1)\,(p-2)\,\frac{a'^2}{a^2}
= - (p+2)\,{\kappa^2_d} \, a^2\,\Lambda_B\, .
\ee
In order to solve Eq.\,(\ref{res}), we perform a 
coordinate transformation $dy=a(z) dz$ which corresponds to the
following 
parametrization of the line element
\eqb
ds^2=a^2(y)\,(-dt^2+ dx_1^2 + dx_2^2 + ... + dx_p^2) + dy^2\ .
\eqe
In these coordinates one gets 
\be
\label{eom}
p\,(p+1)\,\Bigl(\frac{a''}{a}
+\frac{p}{2}\,\frac{a'^2}{a^2}\Bigr)
= -(p+2)\,\kappa_d^2 \,\Lambda_B\, ,
\ee
while Eq.\,(\ref{eq2}) becomes
\eqb
\label{eq2'}
\frac{p}{2}\,(p+1)\,\Bigl(\frac{a'}{a}\Bigr)^2
= \kappa_d^2\left[-\Lambda_B-\frac{(p+1)\alpha}{(aL)^{p+2}}\right]\, ,
\eqe
where now the prime stands for 
differentiation with respect to $y$. 
For $\Lambda_B<0$ the general solution to Eq.\,(\ref{eom}) is
\be
\label{L<0}
a^{(p+2)/2}(y)=A_1\,\cosh(\omega y) + A_2\,\sinh(\omega y)\, ,
\ee
where
\be
\omega^2 =\frac{(p+2)^2}{2p\,(p+1)}\, \kappa_d^2\,|\Lambda_B|\, ,
\label{omega}
\ee
and $A_{1,2}$ are integration constants. For $\Lambda_B>0$ the solution
is 
given by (\ref{L<0}) with cosh and sinh replaced by cos and sin,
respectively. 
Finally, for $\Lambda_B=0$, the solution reads\footnote{In the conformal
frame an exact form of
the solution for $\La_B=0$ was presented in Ref.\,\cite{Garriga}.} 
\be
\label{L=0}
a^{(p+2)/2}(y)=A_1 y + A_2\, .
\ee 
Let us concentrate first on the case $\La_B<0$. The solution (\ref{L<0})
can be rewritten in a more convenient form:
\be
\label{coshst}
a^{(p+2)/2}(y)=a_0^{(p+2)/2}\,\Biggl\{$ \begin{tabular}{l}
$\cosh[\omega (y-y_0)],   \qquad {\rm if} \,\,A_1>A_2$ \\[3mm]
$\sinh[\omega |y-y_0|],   \qquad\,\, {\rm if} \,\,A_1<A_2$
\end{tabular}$
\ee
Inserting these solutions  
into Eq.\,(\ref{eq2'}), we obtain the following consistency conditions
\begin{eqnarray}
\label{consist}
\tanh^2[\omega (y-y_0)]&=&1-\frac{(p+1)\alpha}{|\La_B|\,(a_0L)^d}
\,{\cosh^{-2}[\omega (y-y_0)]}\label{c1} \\[2mm]
\coth^2[\omega (y-y_0)]&=&1-\frac{(p+1)\alpha}{|\La_B|\,(a_0L)^d}
\,{\sinh^{-2}[\omega (y-y_0)]}\label{c2},
\end{eqnarray}
respectively. It follows from Eqs.\,(\ref{c1}) and (\ref{c2}) that 
\eqb
|\La_B|\,L^d=(\pm)\,\frac{(p+1)\alpha}{a_0^{p+2}}\, .
\label{cond1}
\eqe
Eq.\,(\ref{cond1}) demands $\alpha$ to be positive for the cosh-type 
and negative for the sinh-type solution. 
For the 5-dimensional case this was also realized in Ref.~\cite{Muko}. 
Note that the parameter $L=z_2-z_1$, appearing in the above constraints, 
may be written in terms of the coordinates $y_1$ and $y_2$ in 
the non-conformal frame as
\be
\label{ct}
L= \int_{y_1}^{y_2} \fr {dy}{a(y)} = \fr{1}{a_0\omega}I(\omega y_1, 
\omega y_2)\ ,
\label{Lb}
\ee 
where the integral $I$ is defined as follows
\eqb
\label{I}
I(\omega y_1,\omega y_2 )\equiv \int^{\omega y_2 }_{\omega y_1}\fr {d\zeta}
{\cosh^{2/(p+2)}(\zeta-\zeta_0)}\,,
\eqe
where $\zeta_0=\omega y_0$, for the cosh-type solution. A similar expression
(with cosh replaced by sinh) holds for the sinh-type solution. 

Finally, the {\it jump} conditions (\ref{jump}) for the derivative of
the 
scale factor on the branes, expressed in non-conformal coordinates, lead
to the constraints
\be
\label{jc1}
\frac{4\omega}{p+2}\tanh\,[\omega\,(y_k-y_0)]=
(-1)^{k}\frac{\kappa^2_d}{p}\,\La_k\ ,\ \ \ \ \ \ (k=1,2)\ 
\label{conda++}\ee
for the cosh-type solution and
\be
\label{jc2}
\frac{4\omega}{p+2}\coth\,[\omega\,(y_k-y_0)]=
(-1)^{k}\frac{\kappa^2_d}{p}\,\La_k\ ,\ \ \ \ \ \ (k=1,2)\ 
\label{conda+-}\ee
for the sinh-type.
The position of each brane with respect to the point $y=y_0$ determines
the sign of its tension. Note that the
cosh-type solution is characterized by the existence of a minimum
at $y=y_0$ while, at the same point, the sinh-type solution has a
singularity. We always assume that the first brane is located to
the left of $y_0$ ($y_0-y_1>0$), and $\Lambda_1$ is positive.
In order to avoid the singularity in the case of the sinh-type solution 
we place the second brane on the same side of $y_0$. 
According to (\ref{conda+-}), $\Lambda_2$ then should be negative. On the
other hand, in the case of the cosh-type solution the second brane
can be placed to the right of the minimum thus allowing for
a configuration with two positive-tension branes (such a solution
was first studied in Ref.\cite{kop1}). An alternative way of
compactifying the extra dimension arises in this case: discarding the
second brane and identifying the two minima at $y=\pm y_0$, single-brane
models with compact extra dimension may be constructed. The construction
and investigation of the cosmological properties of these models were 
conducted in Refs. \cite{kkop1,kkop2}. 

A remark about the fine-tuning of parameters in our model is in order
at this point. By fixing the position of one of the two branes and imposing
that $a(y_*)=1$, for some $y_*\not=y_0$, we find $y_0$ as a function of
$a_0$. Invoking the consistency condition (\ref{cond1}), $a_0$ is also
fixed in terms of $\Lambda_B$ and the parameter $L$, which is related to
the location of the second brane, and, thus, to the inter-brane distance,
through eq. (\ref{ct}). Before considering the jump conditions, these
two quantities together with the two brane tensions, are free parameters
of the model. The two jump conditions, evaluated at the locations of the
two branes, will fix two of these parameters leaving the other two free.
Here, we choose to fix the bulk cosmological constant and the inter-brane
distance and have as input parameters the two brane tensions. In Section 3,
we will demonstrate\footnote{In Section 3, we will not impose the
condition $a(y_*)=1$. In return, the coordinates $y_1$ and $y_0$ will
conveniently be fixed, in order to simplify our analysis. The result is
the same with both methods: we are left with one integration constant,
$a_0$ (or alternatively $a_+$, see Sec. 3), whose equilibrium value will
be fixed by the consistency condition (\ref{cond1}). The two jump
conditions will fix, in turn, the bulk cosmological constant and the
location of the second brane.} that the
fine-tuning of $\Lambda_B$ and of the location of the second brane
guarantees the existence of static solutions with vanishing $(d-1)$-dimensional
effective cosmological constant and a constant radion field.

Let us investigate the reason that made the construction
of such a geometry possible in the case of the physically more 
interesting $\cosh$-like
solution. In Ref.~\cite{kop1} a simplified approach was pursued:
An unknown mechanism created an effective potential for the 
scale factor of the extra dimension. Varying this contribution with
respect to the metric, 
a $y$-dependent (55)-component of the energy-momentum tensor was found
in addition to the contribution due to the bulk cosmological constant. 
Since the stabilization mechanism was unknown, it was assumed 
in Ref.~\cite{kop1} that the components $T_{00}$ and $T_{ii}$ were not
altered.
In the present work we observe that the 
Casimir stress qualitatively leads to the behaviour of the scale 
factor as in Ref.~\cite{kop1} where 
the existence of a minimum allowed for a configuration 
with two positive-tension branes. Moreover, the $y$-dependent
contribution 
to the total $T_{55}$ given as  
\be
T_{55} = |\Lambda_B| - \frac{(p+1)\,\alpha}{L^d a_0^{p+2} 
\cosh^2[\omega (y-y_0)]}=
|\Lambda_B| - \frac{|\Lambda_B|}{\cosh^2[\omega (y-y_0)]}\,
\label{T55}
\ee
has the same form which the analysis in Ref.~\cite{kop1} demanded.
In addition, the known behaviour of the solution in the bulk allowed us
to derive its contribution to the remaining components of the energy-momentum
tensor and thus to include its effect on the spacetime structure. 
Therefore, the non-isotropic contributions to $T_{00}$ and $T_{ii}$ are
not zero. 
However, they are different from that to $T_{55}$ indicating the 
breaking of the 5-dimensional Poincar{\'e} covariance. 
As a result, the Casimir stress generated by a conformally coupled scalar
field fullfills all the conditions necessary for the existence of a
configuration with two positive-tension branes. The stability of this
configuration, and of the one with a pair of
positive-negative tension branes, under adiabatic perturbations, needs to
be studied.

Concluding this section, let us make a few comments on the solutions 
in the case of positive and zero bulk cosmological constant. For
$\Lambda_B>0$, Eq. (\ref{eq2'}) leads to a constraint similar to
(\ref{cond1}) with $(\pm)$ replaced by $(-)$, for both cos and sin-type 
solutions. Thus, the solution exists only in the case where $\alpha<0$.
Depending on the location of the branes, the system can accommodate
positive-negative,
negative-negative, or even positive-positive tension branes 
due to the oscillating behaviour of the solution. 
The branes of the last configuration are geodesically disconnected by a
pair of singularities. In any case, the hierarchy problem cannot be
solved 
since the warp factor is bounded by a small value. For 
$\Lambda_B=0$, the solution (\ref{L=0}), when
substituted in Eq. (\ref{eq2'}), leads to the result
\be
A_1^2=-\frac{2\kappa^2_d}{p}\,\frac{\alpha}{L^d}\,,
\ee
which again requires $\alpha$ to be negative. 
The {\it jump} conditions then yield
\be
\left[\frac{a(y_2)}{a(y_1)}\right]^{(p+2)/2}=-\frac{\Lambda_1}{\Lambda_2}\,,
\ee
meaning that one of the two branes has negative tension. 
This result could also be inferred from the monotonic behaviour of the
solution (\ref{L=0}). The resolution of the hierarchy seems possible
at first sight. However, the 
fine-tuning of the ratio of the two brane tensions 
introduces a new hierarchy into the model.


\sect{Properties of the radion field}

In this section we perform the derivation of the 4-dimensional effective 
theory following from the higher-dimensional theory (\ref{act}) upon
dimensional reduction. Our main goal is the
determination of a kinetic term and an effective 
potential for the fluctuations of 
proper size of the extra dimension which are 
related to those of the canonically 
normalized radion field. These two quantities decide whether and how
stable 
the static configurations of the previous section are. 
We will consider ``$+$'' single-brane models as well as  ``$++$''
and ``$+-$'' two-brane
models (``$+$'' and ``$-$'' refer to the sign of the brane tensions). 
In what follows we specialize to the case $d=5$.

In order to study the size fluctuations, 
we restore the scale factor $b$ of the extra dimension
and consider it as a four-dimensional field
\eqb
\label{dl}
dy=b(x_{\perp})d\xi\, .
\eqe
Thereby, $\xi$ is a dimensionless 
coordinate and $x_{\perp}\equiv (t,x^1,x^2,x^3)$. 
The static solution corresponds to the equilibrium value 
$\la b \ra =b_0$ which satisfies the constraints listed in 
the previous section. In terms of the coordinate $\xi$ we choose
the position of the first brane to be fixed at the point
$\xi_1=-1$  while the point $y_0$ corresponds to $\xi_0=0$. 
The second brane is located at $\xi=\lambda$ with $\lambda>0$ 
corresponding to $\Lambda_2>0$ for the cosh-type solution.
For the sinh-type solution 
positive values of $\lambda$ are 
excluded due to the singularity at $\xi=0$. The radion field is 
proportional to the deviation of $b$ from its equilibrium value $b_0$,
\be
b(x_{\perp})-b_0\equiv Z^{-1/2} r(x_{\perp})\ , 
\ee
where the factor $Z$ as well as the factor in front of
$(b(x_{\perp})-b_0)^2$ 
can be obtained by an explicit integration of the action (\ref{act})
over  the 
transverse coordinate $\xi$. Fluctuations of $b(x_{\perp})$
about $b_0$ entail $x_{\perp}$-dependent  fluctuations of $a$ \cite{Goldberger}.
Considering only terms up to quadratic order in the fluctuations, 
our final result will contain four dimensional gravity, a kinetic and a
mass term for the radion, 
and the coupling of the radion to gravity. In addition, there will be a 
massless scalar $\phi_{(4)}$, the remnant of the five-dimensional
conformally coupled scalar field $\phi$. The four-dimensional, effective
action takes the following form
\eab
\label{acteff}
S_{\tiny{\mbox{eff}}}=-\int d^dx\,\sqrt{-g_4}\,\left\{- \frac{1}
{2\kappa^2_4}R_{(4)}+\frac{1}{2}\pd_\mu r\,\pd^\mu r
+\frac{1}{2}\pd_\mu\phi_{(4)}\,\pd^\mu\phi_{(4)}+\frac{1}{2}m^2_{r}\,
r^2 + \gamma_{(4)}\phi_{(4)}^2 R_{(4)} + {\cal L}(\phi_{(4)},
\pd_\mu r\,\pd^\mu r) \right\}\ .
\eae
${\cal L}(\phi_{(4)},\pd_\mu r\,\pd^\mu r)$ denotes possible interaction terms
of the radion with the $\phi_{(4)}$ field  which are of no interest 
in this paper. For the stability analysis it is only necessary to
investigate 
the free sector of the effective theory for the radion field. 
We treat cosh- and sinh-type
solutions separately in the remainder of this section.

\subsection{Sinh-type solution}

In order to determine the effective, four-dimensional 
theory for the radion field, we need to start from the 
five-dimensional action and perform the $\xi$-integration explicitly. 
For this purpose we rewrite the dependence of the scale factor $a$ in
terms of 
its value on the first brane, in the case of sinh-type solutions, as
follows 
\be
a(x_{\perp},\xi)= {a_+}(x_{\perp})
\fr{\sinh^{2/5}(\omega b(x_{\perp})|\xi|)}{\sinh^{2/5}
(\omega b(x_{\perp}))}\,,
\label{ansatz1}
\ee
with
\be
a_{+}(x_{\perp}) = a_0\,\sinh^{2/5}(\omega b(x_{\perp}))\,.
\ee
Inserting Eq.\,(\ref{ansatz1}) into Eq.\,(\ref{act}) under consideration of
Eq.\,(\ref{dl}), we obtain the following decomposition of the radion action
\be
\label{acts}
S_r^{\tiny\mbox{eff}} = S_r^{\tiny\mbox{kinetic}} + 
S_r^{\tiny\mbox{interaction}} + S_r^{\tiny\mbox{potential}}\,.
\ee
Thereby, the terms $S_r^{\tiny\mbox{kinetic}}$ and
$S_r^{\tiny\mbox{interaction}}$ 
may be computed together. They include
kinetic terms for $b$ as well as an interaction term with $R_{(4)}$ and
follow from the terms of the 5-dimensional scalar curvature $R_{(5)}$
which involve derivatives of the metric with respect to the 4-dimensional
coordinates. We have 
\be
S_r^{\tiny\mbox{kinetic}}+S_r^{\tiny\mbox{interaction}} =
\int d^4 x \sqrt{-g_{(4)}}\,\biggl\{
A(\omega b)\,R_{(4)}
+B(\omega b)\, \partial_\mu b\, \partial^\mu b \biggr\}\,,
\label{skinb}
\ee
where
\be
A(\omega b) = \frac{b}{\kappa^2_5}\int_{|\lambda|}^{1} d\xi
\,{\sinh^{4/5}(\omega b|\xi|)\over \sinh^{4/5}(\omega b)}\,,
\label{kinetic1}
\ee
\be
B(\omega b) = \fr{12\omega}{5\kappa^2_5}\int_{|\lambda|}^{1} d\xi
\,{\sinh^{4/5}(\omega b|\xi|)\over \sinh^{4/5}(\omega b)}\,
\biggl\{
-[\coth(\omega b)-\xi\coth(\omega b\xi)]
+ \fr {2\omega b}{5}
\Bigl[\coth(\omega b)-\xi\coth(\omega b\xi)\Bigr]^2\biggr\}\,.
\label{kinetic2}
\ee
Eq. (\ref{kinetic1}), upon integration will lead to the definition
of the four-dimensional Newton's constant in terms of the
5-dimensional one and the size of the extra dimension.

The potential part of Eq.\,(\ref{acts}) includes several contributions
due to the brane tensions, the bulk cosmological constant, the Casimir
energy, the terms from the five-dimensional
curvature $R_{(5)}$ involving $\xi$-derivatives of the metric,
and the delta-functional
contributions from the discontinuities of $a'$ at the boundaries
$\xi=-1,~\lambda$. After integrating over $\xi$, 
the expression for $S_r^{\tiny\mbox{potential}}$ reads
\begin{eqnarray}
S_r^{\tiny\mbox{potential}} &=& -\int d^4 x \sqrt{-g_{(4)}}\,V^{\tiny\mbox{eff}}
\nonumber\\[1mm]
&=& -\int d^4 x \sqrt{-g_{(4)}}
\Biggl[ \Lambda_1 - \fr{5}{2}\fr{|\Lambda_B|}{\omega}\coth(\omega b)+
\frac{\sinh^{8/5}(\omega b |\lambda|)}{\sinh^{8/5}(\omega b)}\,\biggl(\Lambda_2
+ \fr{5}{2}\fr{|\Lambda_B|}{\omega}\coth(\omega b|\lambda|)\biggr)
\nonumber\\[1mm]
&~& \hspace*{6.5cm}-\, \frac{1}{\sinh^{8/5}(\omega b)}
\,\biggl(\fr{|\Lambda_B|}{2\omega}\,I(\omega b)
-\fr{2\alpha \omega^4}{I^4(\omega b)}\biggr)\Biggr ],
\label{poten+-}
\end{eqnarray}
where $\sqrt{-g_{(4)}}=a_+^4$. 
Note that the result of integration over $\xi$ can be expressed in terms 
of the integral $I$, defined in eq. (\ref{I}), which now takes the form
\be
I(\omega b)=\int_{\omega b|\lambda|}^{\omega b}\fr {d\zeta}
{\sinh^{2/5}\zeta}\ .
\ee
We may now expand the effective potential up to second order in
$b-b_0$, i.e.
\be
V^{\tiny\mbox{eff}}(b)=V^{\tiny\mbox{eff}}(b_0) + 
\frac{\delta V^{\tiny\mbox{eff}}}{\delta b}\biggl|_{b_0}\,(b-b_0)
+\frac{1}{2}\,\frac{\delta^2 V^{\tiny\mbox{eff}}}{\delta b^2}\biggl|_{b_0}
\,(b-b_0)^2 +...\ .
\ee
With the help of the {\it jump} conditions (\ref{conda+-}), which,
when expressed in terms of $\xi$, can be written as
\be
\Lambda_1=\fr{5}{2}\fr{|\Lambda_B|}{\omega}\coth(\omega b)\,,
\qquad \quad
\Lambda_2=-\fr{5}{2}\fr{|\Lambda_B|}{\omega}\coth(\omega b|\lambda|)\,,
\ee
and the additional constraint (\ref{cond1}), it is not difficult 
to show that both $V^{\tiny\mbox{eff}}\Bigl|_{b_0}$ and 
$\delta V^{\tiny\mbox{eff}}/\delta b \Bigl|_{b_0}$
vanish. Hence, we conclude that the constraint (\ref{cond1}) 
in conjunction with the {\it jump} conditions causes 
the effective four-dimensional cosmological constant to be zero
and, at the same time, extremizes the radion effective potential.
Calculating the second derivative of $V^{\tiny\mbox{eff}}$ with respect to
$b$ and inserting the result in the above expansion, we obtain the following
expression for the effective potential of the radion
\begin{eqnarray}
V^{\tiny\mbox{eff}}=
(b-b_0)^2\fr{24}{5}
\fr{\omega^3}{\kappa^2}\sinh^{-8/5}(\omega b_0)\left[
\fr{1}{I(\omega b_0)}\left(\fr{1}{\sinh^{2/5}(\omega b_0)} -
\fr{|\lambda|}{\sinh^{2/5}(\omega b_0|\lambda|)}\right)^2+\nonumber
\right.\\[1mm]\left.
\fr{2}{5}
\left(
\fr{\coth(\omega b_0)}{\sinh^{2/5}(\omega b_0)}-
\fr{|\lambda|^2\coth(\omega b_0|\lambda|)}
{\sinh^{2/5}(\omega b_0|\lambda|)}
\right)\right].
\label{mass+-}
\end{eqnarray}
Varying $\lambda$ between $-1$ and 0, we realize that the radion mass squared
is not sign definite.  In order to draw some definite conclusions about the
sign of the radion mass  and, thus, the stability of the sinh-type solutions,
we will study separately the cases of nearly flat ($\omega b_0 \ll 1$)
and highly warped ($\omega b_0 \gg 1$) extra dimension.

{\bf (A) Nearly flat extra dimension}. In the limit $\omega b \ll1$,
the hyperbolic functions in Eqs. (\ref{kinetic1})-(\ref{kinetic2})
can be expanded in powers of $\omega b$. Keeping only the dominant
terms and performing the integration, the kinetic part (\ref{skinb})
takes the simplified form
\be
S_r^{\tiny\mbox{kinetic}}+S_r^{\tiny\mbox{interaction}} =
\int d^4 x \sqrt{-g_{(4)}}\,\biggl\{
\frac{R_{(4)}}{2 \kappa^2_4}\,\Bigl(\frac{b}{b_0}\Bigr)
-\frac{8b_0}{171}\,(\omega b_0)\,\Bigl[10-|\lambda|^{9/5}
(19-9 |\lambda|^2)\Bigr]\, \partial_\mu b\, \partial^\mu b \biggr\}\,,
\label{skinb2}
\ee
where
\be
\kappa^2_4=\frac{9 \kappa^2_5}{10 b_0}\,\frac{1}{(1-|\lambda|^{9/5})}\,.
\label{newton1}
\ee
In order to recover Einstein's gravity, we perform the conformal
transformation $g^{(4)}_{\mu\nu}=\Omega(b)\,\hat g^{(4)}_{\mu\nu}=
(b_0/b)\,\hat g^{(4)}_{\mu\nu}$. The
interaction term between $b$ and the 4-dimensional scalar curvature
$R_{(4)}$ will then disappear and an extra contribution to the
kinetic term of $b$ will arise. The final form of the kinetic part,
then, is given by
\be
S_r^{\tiny\mbox{kinetic}}+S_r^{\tiny\mbox{interaction}} =
\int d^4 x \sqrt{-\hat g_{(4)}}\,\biggl\{\frac{\hat R_{(4)}}{2 \kappa^2_4}
- \frac{3}{4\kappa^2_4 b_0^2}\, \partial_\mu b\, \partial^\mu b \biggr\}\,,
\label{skinb3}
\ee
where the subdominant contribution to the kinetic term for $b$
in Eq. (\ref{skinb2}), proportional to $(\omega b_0)$, has been
dropped. Note that, after the conformal transformation and the
derivation of the kinetic term of $b$, we may safely
set $b=b_0$ in all places other than $\partial_\mu b$.
>From the above expression, we may easily read the definition of the
canonically normalized radion field
\be
r(x_\perp)=\sqrt{\frac{3}{2\kappa^2_4 b_0^2}}\,(b-b_0)\,.
\label{radion1}
\ee

We now turn to the form of the effective potential (\ref{mass+-}). In the
same limit, $\omega b_0 \ll 1$, it assumes the form
\be
V^{\tiny\mbox{eff}}= (b-b_0)^2\,\frac{24}{5}\,\frac{(1-|\lambda|^{3/5})}
{\kappa_5^2 b_0^3}\,.
\ee
Under the conformal transformation, the effective potential changes as:
$V^{\tiny\mbox{eff}} \rightarrow \hat V^{\tiny\mbox{eff}} =
\Omega^2(b)\,V^{\tiny\mbox{eff}}$. However, this does not affect either
the existence of the extremum or the sign of the second derivative, since
\be
\frac{\delta \hat V^{\tiny\mbox{eff}}}{\delta b}\Bigr|_{b_0}=
\Omega^2(b)\,\frac{\delta V^{\tiny\mbox{eff}}}{\delta b}\Bigr|_{b_0}+
2 \Omega\,\frac{\delta \Omega(b)}{\delta b}\Bigr|_{b_0}\,
V^{\tiny\mbox{eff}}\Bigr|_{b_0}=0 
\ee
\be
\frac{\delta^2 \hat V^{\tiny\mbox{eff}}}{\delta b^2}\Bigr|_{b_0}=
\Omega^2(b)\,\frac{\delta^2 V^{\tiny\mbox{eff}}}{\delta b^2}\Bigr|_{b_0}+
4 \Omega\,\frac{\delta \Omega(b)}{\delta b}\Bigr|_{b_0}\,
\frac{\delta V^{\tiny\mbox{eff}}}{\delta b}\Bigr|_{b_0}+
2\,\biggl[\Bigl(\frac{\delta \Omega(b)}{\delta b}\Bigr)^2+
\Omega\,\frac{\delta^2 \Omega(b)}{\delta b^2}\biggr]\Bigr|_{b_0}\,
V^{\tiny\mbox{eff}}\Bigr|_{b_0}=
\Omega^2(b)\,\frac{\delta^2 V^{\tiny\mbox{eff}}}{\delta b^2}\Bigr|_{b_0}\,.
\ee
Using the definition of the radion field (\ref{radion1}) and of the
4-dimensional Newton's constant (\ref{newton1}), we find the radion
mass to be
\be
m_r^2=\frac{144}{25 b_0^2}\,\frac{(1-|\lambda|^{3/5})}
{(1-|\lambda|^{9/5})}\,.
\label{mass1}
\ee
>From the above me may conclude that the radion mass remains always
positive in the limit of nearly flat extra dimension, thus, ensuring the
stability of the sinh-type solution, independently of the
position of the negative brane. 
Moreover, it remains well defined for
the whole range of values $-1<\lambda <0$, increasing when
$|\lambda| \rightarrow 0$, as the size of the extra dimension increases,
and decreasing in the opposite limit $|\lambda| \rightarrow 1$.

{\bf (B) Highly warped extra dimension}. In this case we assume that
$\omega b \gg 1$ and write the hyperbolic functions, appearing in the
kinetic and potential parts of the effective action, in terms of
exponentials. Neglecting the exponentially suppressed contributions,
we may then easily perform the corresponding integration over the
extra dimension. Note, however, that our approximation breaks down
when $|\xi| \sim (\omega b)^{-1}$, i.e. when the negative tension
brane is located close to the singularity. We may easily
see, by simple numerical analysis that, as we increase $\omega b_0$,
the positivity of the effective potential
(\ref{mass+-}) demands the negative tension brane to be located
further and further away from the point $\xi=0$. For example, for
$\omega b_0 =20$, the radion mass squared will be negative definite if the
negative tension brane is located in the interval $-0.07 < \lambda < 0$,
while for $\omega b_0 = 86$, relevant to the Randall-Sundrum set-up,
if $-0.09 < \lambda <0$. Therefore, in order to ensure the stability
of our sinh-type solutions in the case of large warping, we need to
place the second brane away from the singularity. In this case, the
area around $\xi \simeq 0$ is excluded and our assumption that
$\omega b |\xi| \gg 1$ will always hold for the stable solutions.

The kinetic and interaction terms in this case take the form
\be
S_r^{\tiny\mbox{kinetic}}+S_r^{\tiny\mbox{interaction}} =
\int d^4 x \sqrt{-g_{(4)}}\,\biggl\{
\frac{R_{(4)}}{2 \kappa^2_4}\,\Bigl(1-e^{4\omega b_0 (|\lambda|-1)/5}\Bigr)
-\frac{6\omega}{5 \kappa_5^2}\,(|\lambda|-1)^2\,
e^{4\omega b (|\lambda|-1)/5}\, \partial_\mu b\, \partial^\mu b \biggr\}\,,
\label{skinb4}
\ee
where
\be
\kappa^2_4=\frac{2 \omega}{5}\,\kappa^2_5\,.
\label{newton2}
\ee
Once again, we need to perform a conformal transformation, of the form
$g^{(4)}_{\mu\nu}=\Bigl(1-e^{4\omega b (|\lambda|-1)/5}\Bigr)^{-1}
\hat g^{(4)}_{\mu\nu}$, in order to recover Einstein's gravity.
Then, we arrive at the result
\be
S_r^{\tiny\mbox{kinetic}}+S_r^{\tiny\mbox{interaction}} =
\int d^4 x \sqrt{-\hat g_{(4)}}\,\biggl\{\frac{\hat R_{(4)}}{2 \kappa^2_4}
-\frac{6\omega}{5 \kappa_5^2}\,(|\lambda|-1)^2
e^{4\omega b_0 (|\lambda|-1)/5}\, \partial_\mu (b-b_0)\,
\partial^\mu (b-b_0) \biggr\}\,,
\label{skinb5}
\ee
where the subdominant contributions to the kinetic term, after the
conformal transformation and the expansion around $b_0$, have been
ignored.  In this case, the canonically normalized radion field
is defined as
\be
r(x_\perp)=\sqrt{\frac{12 \omega}{5\kappa^2_5}\,(|\lambda|-1)^2}\,
e^{2\omega b_0 (|\lambda|-1)/5}\,(b-b_0)\,.
\label{radion2}
\ee

The effective potential (\ref{mass+-}), in the limit $\omega b_0 \gg 1$,
reads
\be
\hat V^{\tiny\mbox{eff}}= (b-b_0)^2\,\frac{192\,\omega^3}{25\,\kappa^2_5}\,
\frac{(1-|\lambda|)^2}{(1-e^{-4\omega b_0 (1-|\lambda|)/5})^2}
\,\frac{e^{-2\omega b_0}}{(1-e^{-2 \omega b_0(1-|\lambda|)/5})}\,,
\label{veff-final}
\ee
which, when expressed in terms of the radion field, it yields the
corresponding radion mass
\be
m_r^2=\frac{32}{5}\,\omega^2\,\Bigl(\frac{M_W}{M_P}\Bigr)^3\,
e^{-2\omega b_0 |\lambda|}\,.
\label{mass2}
\ee
In the above expression, we have used the geometrical 
explanation of the Planck scale/weak scale hierarchy which in the limit of
large warping is equivalent to
\be
\frac{a_-}{a_+} \simeq e^{-2 \omega b_0 (1-|\lambda|)/5}=\frac{M_W}{M_P}
\ee
and ignored the exponentially suppressed terms in the denominator
of Eq. (\ref{veff-final}).
As expected, since we restricted ourselves to the study of the stable
sinh-type solutions, the radion mass squared turned out to be positive
and proportional to the ratio $M_W^3/M_P$, since $\omega^2 \sim M_P^2$.
However, there is an extra suppression factor that decreases very
rapidly as $\omega b_0 \rightarrow \infty$ leading to an additional strong
suppression of the radion mass. This factor depends on the location
of the negative tension brane, leading to a larger suppression
when the second brane is located close to the first one and to a
smaller suppression when the second brane is located closer to the singularity.
The relative change however is very small since $|\lambda|= {\cal O}(1)$.

\subsection{Cosh-type solutions}

For positive $\alpha$ (cosh-type solution) we may write
the scale factor as follows
\begin{eqnarray}
\label{ansatz2}
a(x_{\perp},\xi)={a_+}(x_{\perp})
\fr{\cosh^{2/5}(\omega b(x_{\perp})\xi)}{\cosh^{2/5}
(\omega b(x_{\perp}))}\,,
\end{eqnarray}
where, similarly to the previous subsection, we have defined
\be
a_+(x_{\perp}) = a_0\cosh^{2/5}(\omega b(x_{\perp}))\,.
\ee
We first focus on the kinetic and interaction terms. Substituting the
above ansatz in the 5-dimensional action and decomposing $R_{(5)}$,
we obtain
\be
\label{Sr}
S_r^{\tiny\mbox{kinetic}} + 
S_r^{\tiny\mbox{interaction}}=\int d^4 x \sqrt{-g_{(4)}}\,\biggl\{
\tilde A(\omega b)\,R_{(4)}
+\tilde B(\omega b)\, \partial_\mu b\, \partial^\mu b \biggr\}\,,
\ee
where now
\be
\tilde A(\omega b) = \fr{b}{\kappa^2_5}\int_{-1}^{\lambda} d\xi
\,{\cosh^{4/5}(\omega b\xi)\over \cosh^{4/5}(\omega b)}\,,
\label{kinetic3}
\ee
\be
\tilde B(\omega b) = \fr{12\omega}{5\kappa^2_5}\int_{-1}^{\lambda} d\xi
\,{\cosh^{4/5}(\omega b\xi)\over \cosh^{4/5}(\omega b)}\,
\biggl\{
-[\tanh(\omega b)-\xi\tanh(\omega b\xi)]
+ \fr {2\omega b}{5}
\Bigl[\tanh(\omega b)-\xi\tanh(\omega b\xi)\Bigr]^2\biggr\}\,.
\label{kinetic4}
\ee

In analogy to the previous subsection we obtain for the potential part
the expression
\begin{eqnarray}
S_r^{\tiny\mbox{potential}} &=& -\int d^4 x \sqrt{-g_{(4)}}\,V^{\tiny\mbox{eff}}
\nonumber\\[1mm]
&=& -\int d^4 x \sqrt{-g_{(4)}}
\Biggl[ \Lambda_1 - \fr{5}{2}\fr{|\Lambda_B|}{\omega}\tanh(\omega b)+
\frac{\cosh^{8/5}(\omega b \lambda)}{\cosh^{8/5}(\omega b)}\,\biggl(\Lambda_2
- \fr{5}{2}\fr{|\Lambda_B|}{\omega}\tanh(\omega b\lambda)\biggr)
\nonumber\\[1mm]
&~& \hspace*{6.5cm}+\, \frac{1}{\cosh^{8/5}(\omega b)}
\,\biggl(\fr{|\Lambda_B|}{2\omega}\,I(\omega b)
-\fr{2\alpha \omega^4}{I^4(\omega b)}\biggr)\Biggr ],
\label{poten}
\end{eqnarray}
where
\be
I(\omega b)=\int^{\omega b\lambda }_{-\omega b}\fr {d\zeta}
{\cosh^{2/5}\zeta}\,.
\ee
Using the {\it jump} conditions (\ref{conda++}), which can be
written in the simple form
\be
\Lambda_1=\fr{5}{2}\fr{|\Lambda_B|}{\omega}\tanh(\omega b)\,,
\qquad \quad
\Lambda_2=\fr{5}{2}\fr{|\Lambda_B|}{\omega}\tanh(\omega b\lambda)\,,
\ee
we once again see that both $V^{\tiny\mbox{eff}}$ and
$\delta V^{\tiny\mbox{eff}}/\delta b$ vanish when evaluated at the static
solution. Performing the expansion around $b_0$, we finally get the following
expression for the effective potential
\begin{eqnarray}
V^{\tiny\mbox{eff}}= -(b-b_0)^2\fr{24}{5}
\fr{\omega^3}{\kappa^2_5}\cosh^{-8/5}(\omega b_0)\left[
\fr{1}{I(\omega b_0)}\left(\fr{1}{\cosh^{2/5}(\omega b_0)} + 
\fr{\lambda}{\cosh^{2/5}(\omega b_0\lambda)}\right)^2+\nonumber
\right.\\[1mm]\left.
\fr{2}{5}
\left(
\fr{\tanh(\omega b_0)}{\cosh^{2/5}(\omega b_0)}+
\fr{\lambda^2\tanh(\omega b_0\lambda)}
{\cosh^{2/5}(\omega b_0\lambda)}
\right)\right].
\label{mass++}
\end{eqnarray}
This is negative definite for any value of $b\neq b_0$. As a result, we conclude
that a positive $\alpha$ leads to a negative $m_r^2$ and thus to a 
tachyonic potential for the radion field.  Although the static configuration
with two positive tension branes, presented in the previous section, corresponds
to an extremum of the effective potential this extremum is a maximum. 
Therefore, any small perturbation of this static solution induces an instability
causing the radion field to run away from its equilibrium value. Hence, there
is no stabilization of the  extra dimension.

For completeness, we now  briefly give the results for the expression of the
radion field, in terms of $(b-b_0)$, and of the radion mass in the two
cases of nearly flat ($\omega b_0 \ll 1$) and highly warped ($\omega b_0 
\gg 1$) extra dimension.

{\bf (A) Nearly flat extra dimension}. In the limit $\omega b \ll 1$,
the integration in Eq. (\ref{kinetic3}) gives rise to an interaction term
between $b$ and $R_{(4)}$, as in the case of the sinh-type solution. The
same conformal transformation,
$g^{(4)}_{\mu\nu}=(b_0/b)\,\hat g^{(4)}_{\mu\nu}$,
removes this term and gives the dominant contribution to the kinetic term
of $b$. Then, the kinetic part of the effective action reduces again to
Eq. (\ref{skinb3}), leading to the same definition (\ref{radion1}) for the
radion field, while
\be
\kappa^2_4=\frac{\kappa^2_5}{2 b_0 (\lambda+1)}\,.
\ee
In the same limit, the effective potential (\ref{mass++}) assumes the form
\be
\hat V^{\tiny\mbox{eff}}= -(b-b_0)^2\,\frac{24 \omega^2}{5\kappa_5^2 b_0}\,
(\lambda+1)=\frac{1}{2}\,\Bigr(-\frac{16 \omega^2}{5}\Bigr) r^2(x_\perp)\,,
\ee
from which we may easily read the expression of the radion mass squared
which is indeed negative. However, in the limit $\omega \rightarrow 0$,
this mass goes to zero leaving behind a static cosh-type solution
which is a saddle point of the effective action instead of a maximum.
Nevertheless, this semi-stable solution is plagued
by the existence of a massless radion field.

{\bf (B) Highly warped extra dimension}. In this case, the analysis is similar
to the one of the previous subsection, in the limit $\omega b_0 \gg 1$. Upon
integration of Eqs. (\ref{kinetic3})-(\ref{kinetic4}), we find the result
\be
S_r^{\tiny\mbox{kinetic}}+S_r^{\tiny\mbox{interaction}} =
\int d^4 x \sqrt{-g_{(4)}}\,\biggl\{
\frac{R_{(4)}}{2 \kappa^2_4}\,\Bigl(e^{4\omega b (\lambda-1)/5}
-e^{-8\omega b/5}\Bigr) +\frac{6\omega}{5 \kappa_5^2}\,
\Bigl[(\lambda-1)^2 e^{4\omega b (\lambda-1)/5} -4 e^{-8 \omega b/5}\Bigr]\,
\partial_\mu b\, \partial^\mu b \biggr\}\,,
\ee
where $\kappa^2_4$ is defined as in Eq. (\ref{newton2}). The
conformal transformation
$g^{(4)}_{\mu\nu}=\Bigl(e^{4\omega b (\lambda-1)/5}-e^{-8 \omega b/5}\Bigr)^{-1}
\hat g^{(4)}_{\mu\nu}$ restores Einstein's gravity and leads to a canonically
normalized kinetic term for the radion field, which is given
by Eq. (\ref{radion2})
with $(|\lambda|-1)$ replaced by $(\lambda+1)$. The effective potential
(\ref{mass++}), in the limit $\omega b \gg 1$, reads
\be
\hat V^{\tiny\mbox{eff}}= -(b-b_0)^2\,\frac{192\,\omega^3}{25\,\kappa^2_5}\,
e^{-2\omega b_0}\,
\frac{e^{-4\omega b_0/5}(1+ \lambda e^{-2 \omega b_0
(\lambda-1)/5})^2 + (1+ \lambda^2 e^{-2 \omega b_0(\lambda-1)/5})}
{(e^{4\omega b (\lambda-1)/5}-e^{-8 \omega b/5})^2}\,,
\label{veff-final2}
\ee
which yields the radion mass squared
\be
m_r^2=-\frac{32}{5}\,\frac{\omega^2}{(\lambda+1)^2}\,e^{-2\omega b_0 \lambda/5}
\Biggl\{\begin{tabular}{r} $\lambda^2\,\Bigl(
\frac{\textstyle a_1}{\textstyle a_2}\Bigr)^{2} \quad
(\lambda \leq 1)$\\[4mm] $\Bigl(\frac{\textstyle a_1}{\textstyle a_2}\Bigr)
\quad \,\,\, (\lambda > 1) $\end{tabular}\,,
\ee
where $(a_1/a_2) \simeq e^{2 \omega b_0 (1-\lambda)/5}$ is the ratio of the
warp factors of the two branes located at $\xi=-1, \lambda$, respectively.
The radion mass squared is always negative definite, as expected. Its 
value strongly depends on the position of the second positive brane and
may vary significantly for various values of $\lambda$. 

As mentioned in the previous section, the cosh-type solution allows
for a single positive-tension brane configuration with compact extra dimension.
The above analysis applies also in that case. Setting $\Lambda_2=0$ and
taking the limit $\lambda \rightarrow 0$, i.e. moving the ``empty''
brane to the location of the minimum in all the above expressions,
one obtains the corresponding results for the single-brane
configuration. It is obvious that the effective potential (\ref{mass++})
maintains its negative sign which also indicates the instability of the ``+''
brane configuration.

\section{Conclusions}

It is often thought that staticity of gravity in a spacetime with a
compact, 
extra dimension {\em requires} the introduction of two branes possessing
tensions of 
opposite sign. This statement is usually 
illustrated using an analogue in electrostatics. The presence of one 
positive-tension brane creates a gravitational field with the arrows of 
the field strength lines directed away from the brane. Since the
dimension 
is compact these lines should meet at a point associated with a 
negative-tension brane. This argument is flawed since it is 
possible that at one point the gravitational field 
strength simply becomes zero, thus allowing to complete the circle 
{\em without} introducing a negative tension object. This idea was
advocated 
in \cite{kkop1,kkop2}, and the concrete realization of such a
compactification 
due to the Casimir stress was discussed in \cite{Muko} and in this paper.

Interestingly enough, a classical force created by a scalar field in the 
bulk does not yield a static situation without introducing a 
metric singularity or a compensating negative-tension brane as it was
shown in 
\cite{kop1,Wisc}. The presence of the quantum effects (Casimir stress),
however, 
does allow for single and two-brane configurations with
only positive tensions as we showed in section 2. The deeper reason for
this is the violation of the classical energy condition by the Casimir 
stress \cite{Muko} since the pressure in the extra dimension exceeds the 
energy density in the bulk. Positive-tension brane configurations
arise in the case of an attractive Casimir force, whereas a repulsive 
Casimir force 
implies static pairs of positive-negative tension branes. 

In section 2 we found exact static solutions to the Einstein's equations, 
in a $d=p+1+1$--dimensional spacetime, in the presence of a bulk
cosmological constant being positive, negative or zero and 
of the Casimir stress 
created by a conformally coupled scalar field. It was shown that 
time-dependent cosmological solutions in a $d$-dimensional,
conformally-flat spacetime are excluded if the Casimir stress is
implemented in the theory. Ignoring the Casimir stress, non-static
solutions, which are generalizations of the RS solution in de Sitter and anti
de Sitter spacetime, were derived. These solutions allowed only for pairs
of positive-negative tension branes with extremely fine-tuned 
brane-tensions. Restoring the Casimir stress and considering 
a negative bulk cosmological constant, we
found two different types of static solutions for the scale factor in the
bulk, namely a cosh-type solution for an attractive Casimir force
($\alpha>0$) and a sinh-type solution for a repulsive Casimir force
($\alpha<0$). The monotonic behaviour of the sinh-type solution was
consistent only
with the presence of a positive-negative tension pair of branes. On the
other hand, the existence of a minimum in the behaviour of the cosh-type
solution allowed for the introduction of two positive tension branes 
or even for the compactification of the extra spatial dimension with 
a single positive-tension brane. These types of
configurations were studied before in Refs.\,\cite{kkop1,kkop2,kop1,kop2},
by assuming the existence of an unknown mechanism to create 
an isotropy breaking contribution to the (55)-component of the 
energy-momentum tensor. 

In the above context we addressed the issue of the stabilization of the
extra dimension. 
Up to quadratic fluctuations about the static solutions and for all 
possible brane configurations, we derived the dimensionally reduced
effective 
theory for the radion field. We calculated the mass squared 
of the radion field in each case. According to our
results, there is indeed stabilization of the positive-negative pair of
branes arising for a repulsive Casimir force and for the branes 
situated in the exponential regime of the 
sinh-solution. This implies that the corresponding static 
solution corresponds to a minimum of the radion 
effective potential. Interestingly, the situation becomes unstable if 
the negative-tension brane is placed close to the singularity. 
However, in the case of stabilization the 
hierarchy between the scales on the two branes produces a
radion mass in the KeV range or smaller which makes this possibility
phenomenologically unacceptable. Similar results demonstrating the
stability of a positive-negative brane configuration in the case of a
repulsive Casimir force, which was also plagued by an unnaturally
small radion mass, were presented by Garriga et al. \cite{Garriga}.
The possibility of stabilizing a similar brane configuration by taking into
account quantum fluctuations of bulk fields was also studied by
Goldberger and Rothstein \cite{more} and the same inconsistency
between the resolution of the hierarchy problem and the natural
stabilization of the inter-brane distance was again pointed out.
The role of the singularity in the stability of the system, when
the second brane is placed close to it, became apparent only in
the framework of our analysis, where exact solutions for the
spacetime background, in the presence of the Casimir force,
were derived. This feature was not
revealed in any of the above works, where the quantum effects
were considered as small perturbations in a fixed background.
On the other hand, the static cosh-type
solution arising for an attractive Casimir force, turned out to be
unstable independently of the position of the two positive tension
branes. This is in agreement with the results by Nam \cite{more}
where a purely attractive Casimir force due to a scalar field fails to
stabilize the radion field. The single positive-tension brane configuration,
that also arises in this case, shares the same kind of instability. As
a result, the Casimir force fails to stabilize any configuration involving
solely positive-tension branes, thus, leaving this question open
for future study.

\section*{Acknowledgments}

The work of R.H. was funded by Deutscher 
Akademischer Austauschdienst (DAAD). This work was supported in part
by the Department of Energy under Grant No. DE-FG-02-94-ER-40823
at the University of Minnesota.
P.K. would also like to acknowledge financial support by EC under
the TMR contract No. HPRN-CT-2000-00148 during the late stages of 
this work.

\bibliographystyle{prsty}

\end{document}